\documentclass{emulateapj}

\newcommand{\kms}{km~s$^{-1}$}
\newcommand{\hal}{H$\alpha$}
\newcommand{\hb}{H$\beta$}
\newcommand{\ov}{SN~2006ov}
\newcommand{\my}{SN~2006my}
\newcommand{\aaa}{SN~2007aa}

\newcommand{\ebv}{$E(B-V)$}
\newcommand{\tp}{$t_p$}
\newcommand{\qrsp}{$q_{\rm{RSP}}$}
\newcommand{\ursp}{$u_{\rm{RSP}}$}
\newcommand{\citepeg}[1]{\citep[{e.g.,}][]{#1}}

\begin{document}

\shorttitle{Aspherical Cores in Three SNe IIP}
\shortauthors{Chornock et al.}

\title{Large Late-time Asphericities in Three Type IIP Supernovae}

\author{Ryan Chornock\altaffilmark{1,2},
Alexei V. Filippenko\altaffilmark{1},
Weidong Li\altaffilmark{1},
and Jeffrey M. Silverman\altaffilmark{1}
}

\altaffiltext{1}{Department of Astronomy, University of California,
                 Berkeley, CA 94720-3411, USA.}
\altaffiltext{2}{Current address: Harvard-Smithsonian Center for
  Astrophysics, 60 Garden Street, Cambridge, MA 02138, USA, 
                 \texttt{rchornock@cfa.harvard.edu}.}

\begin{abstract}
Type II-plateau supernovae (SNe~IIP) are the results of the explosions
of red supergiants and are the most common subclass of core-collapse
supernovae.  Past observations have shown that the outer
layers of the ejecta of SNe IIP are largely spherical, but the degree
of asphericity increases toward the core.  We present evidence for
high degrees of asphericity in the inner cores of three recent
SNe~IIP (SNe~2006my, 2006ov, and 2007aa), as revealed by late-time
optical spectropolarimetry.  The three objects were all selected to
have very low interstellar polarization (ISP), which minimizes the
uncertainties in ISP removal and allows us to use the continuum
polarization as a tracer of asphericity.  The three objects have
intrinsic continuum polarizations in the range of $0.83-1.56$\% in
observations taken after the end of the photometric plateau, with the
polarization dropping to almost zero at the wavelengths of strong
emission lines.  Our observations of \aaa\ at earlier times, taken on
the photometric plateau, show contrastingly smaller continuum
polarizations ($\sim0.1$\%).  The late-time \hal\ and [\ion{O}{1}]
line profiles of \ov\ provide further evidence for asphericities in
the inner ejecta.  Such high core polarizations in very ordinary
core-collapse supernovae provide further evidence that essentially all
core-collapse supernova explosions are highly aspherical, even if the
outer parts of the ejecta show only small deviations from spherical
symmetry. 
\end{abstract}


\section{Introduction}

Core-collapse supernovae (SNe) are the explosive deaths of massive
stars.  Indirect inferences that these energetic explosions are
aspherical have come from such evidence as the high space velocities
of the pulsars produced by the SNe \citep{ll94} and the structure of
Galactic supernova remnants (e.g., Fesen 2001).  A direct
observational test for the presence of asphericity in SNe was
proposed by \citet{ss82}, who pointed out that the polarizing effects
of electron scattering in a supernova atmosphere would not cancel if
the supernova were aspherical and spatially unresolved by a distant
observer.

The first published detections of supernova polarization were for SN
1987A \citep{bar88,cropper88,mendez88}.  Subsequent observations of
additional SNe led to the discovery that
all core-collapse supernovae with data of sufficient quality exhibit 
polarization, as recently reviewed by \citet{ww08}.  An important test
of our understanding was provided by the late-time \emph{Hubble Space
Telescope} (\emph{HST}) images of SN~1987A which spatially resolve the
ejecta and show that they are elongated along a position angle (P.A.)
consistent with that inferred from the polarimetry \citep{wa02}. 

Our study here focuses on SNe IIP, which represent the single most
common outcome for the ends of the lives of massive stars.
Approximately 50\% of all core-collapse SNe are of this subtype (Li et
al., in prep.; see also Smartt et al. 2009).  Their 
progenitors are relatively low-mass red supergiants in the mass range
of $8.5-16.5$~M$_{\odot}$ \citep{smartt09} that are either isolated or
sufficiently separated from their binary companions so as to leave
their evolution and outer envelopes unaffected
\citep{heger03,li06,smartt09}.  In many ways, their progenitors 
and explosions should be the simplest ones to understand and model.

Despite this, published spectropolarimetry of SNe~IIP is surprisingly
limited.  Most of the objects of this type listed in the compilation
of \citet{ww08} as having available polarimetry are either unpublished
or have only photometric polarimetry.  Our focus here is on
measuring continuum polarization and separating intrinsic continuum
polarization from the effects of interstellar polarization (ISP); with
only photometric data, this is extremely difficult, if not impossible.

The two SNe IIP that have been the most well-studied
spectropolarimetrically are SNe~1999em and 2004dj.  SN~1999em showed
low polarization shortly after explosion, indicating that the outer
ejecta were nearly spherical.  The level of continuum
polarization slowly increased with time in a manner consistent with a
single axis of symmetry while simultaneously the polarization
signatures in the lines grew \citep{le01,ww08}.  SN~2004dj showed no
detectable polarization (after correction for ISP) in the continuum or
lines while on the plateau.  At the end of the plateau stage, the
continuum 
polarization jumped up to 0.56\%, revealing an aspherical core
\citep{doug04dj}.  The polarization percentage slowly decreased
thereafter in a manner attributed to geometric dilution of the
scattering electrons as the ejecta expanded.  The observations of
SN~2004dj prompted the present study in an attempt to determine
whether the results for SN~2004dj were common to SNe~IIP.

Three recent SNe~IIP in the Virgo Cluster have provided an opportunity
to study this class of objects in more detail.  \my\ was
discovered by K.~Itagaki in NGC~4651 on 2006 November 8.82 (UT dates are
used throughout this paper; Nakano \& Itagaki 2006a).  The same
observer also discovered \ov\ in M61 on 2006 November 24.86
\citep{nak06b}.  Both objects were soon spectroscopically confirmed
as SNe~IIP a few months past explosion (Stanishev \& Nielsen 2006;
Blondin et al. 2006).  \aaa\ in NGC~4030 was discovered by T.~Doi on
2007 February 18.308 and spectra obtained by N.~Morrell on Feb. 19.24
showed that it was a SN~IIP similar to SN~1999em at about 20 days
after explosion \citep{doi07,fol07}.
\citet{li07} presented photometric and spectroscopic evidence that both
\my\ and \ov\ were normal Type~IIP SNe discovered near the end of the
photometric plateau stage.

While \citet{li07} identified possible progenitors for SNe~2006my and
2006ov in pre-explosion $HST$ images, subsequent work has rejected those
identifications \citep{doug09,smartt09,crockett09}.  Instead,
\citet{smartt09} and \citet{crockett09}
give upper limits for the zero-age main-sequence 
masses of any progenitor in the \emph{HST} images of 13, 10, and
12 M$_{\odot}$ for SNe 2006my, 2006ov, and 2007aa, respectively.
\citet{smartt09} also quote values for the $^{56}$Ni masses produced
by \ov\ and \my\ of $0.003 \pm 0.002$~M$_{\odot}$ and $0.03 \pm
0.015$~M$_{\odot}$, respectively, based on analyses of the late-time
light curves.  The $^{56}$Ni mass inferred for \ov\ is extremely low, 
comparable to that of the underluminous SN~1999br
\citep{hamuy03,pastorello04}, despite \ov\ having a normal luminosity
on the plateau.

In this paper, we present spectropolarimetry of the above three
SNe~IIP obtained after the end of the photometric plateau.  In
\S2, we describe the observations and corrections for
small amounts of ISP from the Galaxy.  We present the evidence for 
very high intrinsic continuum polarizations in these objects in
\S3 and discuss the interpretation and implications of
our results in \S4.

\section{Observations\label{IIP_obs}}

Unfiltered images of all three SNe in our sample were obtained as
part of our regular monitoring of their host galaxies for 
the Lick Observatory Supernova Search using the 0.76-m Katzman
Automatic Imaging Telescope (KAIT; Filippenko et
al. 2001), although we artificially increased the frequency of
observations from the default search strategy to better sample the
light curves of these objects.  Portions of the \ov\ and \my\ light
curves have already been presented and the reductions described by
\citet{li07}.  Briefly, a point-spread 
function fitting technique was used to subtract pre-SN template images
from the SN observations and aperture photometry was performed on the
subtracted images using DAOPHOT \citep{stetson87} in
IRAF\footnote{IRAF is distributed by the National Optical Astronomy 
Observatories (NOAO), which are operated by the Association of
Universities for Research in Astronomy, Inc., under cooperative
agreement with the National Science Foundation.}.  KAIT unfiltered
observations approximate the standard $R$ band \citep{ligrb}, so the
SN observations were then calibrated to an average of USNO-B1 stars in
the field (up to 20 per field), resulting in systematic
uncertainties in the zero points 
for each field of $0.05-0.18$~mag.  However, only the shapes of the
light curves are relevant for this study. 

We obtained spectopolarimetry of the three SNe using both the Kast
spectrograph on the Lick 3-m Shane telescope \citep{ms93} and the Low 
Resolution Imaging Spectrometer (LRIS) mounted on the Keck I 10-m
telescope \citep{oke95}.  The Kast data were obtained using the
red camera only, with the 300/7500 grating tilted to cover a
wavelength range of approximately $4400-9900$~\AA\ in conjunction with
a GG455 order-blocking filter.  A small 
amount of second-order light is likely present at wavelengths greater
than $\sim9000$~\AA, but none of our conclusions
depend on data at those wavelengths.  A 3$\arcsec$--wide slit gave a
spectral resolution of $\sim16$~\AA.

The LRIS data were taken with both halves of the spectrograph.  All
observations used an identical setup with the D560 dichroic
beam-splitter to split the spectrum near $5500-5600$~\AA, along with
the 600/5000 grism on the blue side and the 400/7500 grating on the
red side.  We used a $1\farcs5$--wide slit to achieve spectral
resolutions of $\sim6$~\AA\ and 9~\AA\ on the blue and red sides,
respectively.  The total spectral coverage was $3170-9240$~\AA, but
the polarization data plotted in this paper are typically truncated
below $\sim4000$~\AA\ because the low blue flux in SNe IIP due to line
blanketing, and decreasing detector sensitivity at short wavelengths,
combine to make the polarimetry extremely noisy at the blue end.  In
addition, this spectral region suffers from 
background-subtraction uncertainties due to the underlying blue
starlight from the host galaxy.  Further details of the individual
observations are given in Table~\ref{IIPobstab}.

\begin{deluxetable*}{lccccccl}
\tablecaption{Details of Observations}
\tablehead{\colhead{Object} & \colhead{UT Date} & 
  \colhead{Age\tablenotemark{a}} & 
  \colhead{Telescope\tablenotemark{b}} &
  \colhead{Total Exposure Time} & \colhead{Seeing} & \colhead{P.A.} &
  \colhead{Notes} \\
 & (YYYY MM DD.DD) & (d) & & (s) & ($\arcsec$) & ($\degr$) &  }
\startdata
\ov\ & 2006 12 20.51 & $-5$ & L3 & 4800 & 2.2 & 144.5 & clouds \\
\ov\ & 2006 12 25.63 & 0 & K1 & 4000 & 1.4 & 125 & clear \\
\ov\ & 2007 01 21.56 & 27 & K1 & 4400 & 0.8 & 125 & clear \\
\my\ & 2007 01 21.50 & 38 & K1 & 3600 & 0.9 & 282 & clear \\
\aaa\ & 2007 03 18.45 & $-50$ & K1 & 3600 & 1.0 & 359.7 & clear, poor guiding \\
\aaa\ & 2007 04 15.48 & $-22$ & K1 & 2800 & 1.8 & 57.5 & clear \\
\aaa\ & 2007 05 10.31 & 3 & L3 & 7200 & 2.1 & 203 & clear \\
\label{IIPobstab}
\enddata
\tablenotetext{a}{Age relative to the end of the plateau, \tp\ (see
  \S3 for a definition).} 
\tablenotetext{b}{L3 = Lick 3-m + Kast spectrograph; K1 = Keck I 10-m
  + LRISp.}
\end{deluxetable*}

Basic two-dimensional image processing and spectral extraction were
accomplished using standard 
tasks in IRAF.  Flux calibration and removal of
atmospheric absorption bands \citep{wh88} was performed using our own
IDL tasks \citep{ma00}.  The individual one-dimensional SN spectra
were combined on a pixel-by-pixel basis, weighted using the error
estimates from the optimal spectral extraction technique
\citep{horne86}, and the final blue-side and red-side spectra (for
LRIS) were scaled and combined across the overlap region.  In those
cases where differential loss of blue light was apparent because the
observed P.A. of the observations deviated from the 
parallactic angle \citep{fi82}, the combined observation was warped to
the shape of the observation obtained closest to the parallactic
angle using a low-order polynomial fit to the ratio of the spectra.

We also extracted some of the emission from nearby \ion{H}{2} regions
along the slits of selected observations.  We fit Gaussians to the
nebular emission lines and obtained recession velocities of 920, 1660,
and 1650~\kms\ for the \ion{H}{2} regions near \my, \ov, and \aaa,
respectively.  These values are within a typical value for a
galactic rotation velocity ($100-200$~\kms) of those listed for the
host-galaxy nuclei in the NASA Extragalactic Database (NED), so we
adopt these redshifts in this paper and have removed them from all
plots.  All references to wavelengths are in the rest frame of each
object.

\subsection{Spectropolarimetry}

The spectropolarimetric observations were reduced following the
standard method of \citet{mrg88}, as implemented by \citet{le01}.  The
angle zero points of the waveplates were determined using
polarimetric standards from \citet{sch92}.  Typically, two
standards were observed per night to confirm the zero point.  In
addition, we noted that the zero points for each instrument were
generally stable over time with small shifts.  Low-polarization
standards from the literature \citep{mat70,ct90,sch92,berd95} were
observed on each night to have negligible polarizations, setting a
limit on the instrumental polarization of $P < 0.1\%$.

This study hinges on our ability to accurately measure the continuum
polarization of SNe.  ISP in both our own Galaxy and the host galaxies
of these SNe can mimic the presence of continuum polarization and
contaminate our results.  All three SNe selected for this study are at
high Galactic latitudes ($b > 59\arcdeg$) and hence low Galactic ISP
is expected \citep{ser75} based on the low columns
(\ebv\ $<0.03$~mag) measured along these lines of sight \citep{sfd98}.
We can directly measure the Galactic contribution to the ISP by
following the lead of \citet{tr95} and selecting stars of low
intrinsic polarizations and sufficiently large distances (estimated
via spectroscopic parallaxes) to place them out of the Galactic
plane.  Such probe stars should sample most of the Galactic
contribution to the ISP along these lines of sight. The probe stars for 
\my\ were BD +17$^\circ$2526 and BD +16$^\circ$2409, those for \ov\ were 
BD +05$^\circ$2618 and BD +04$^\circ$2608, while BD +00$^\circ$2876 and 
BD $-00^\circ$2514 were selected for \aaa.

The probe stars were observed immediately after the object on one
observation date per SN with the slit kept at the same P.A.  Each pair
of stars gave consistent results (within 0.1\% in each Stokes
parameter), so the two stars were averaged to form our estimate of the
Galactic ISP for each SN.  As expected, the Galactic ISPs were very
small, with $V$-band polarizations of 0.07\%, 0.06\%, and 0.16\%
measured for the \my, \ov, and \aaa\ probes, respectively.  For the
sake of completeness, we subtracted low-order polynomial fits to the
Stokes $q$ and $u$ of the combined probes from the Stokes parameters
measured during each SN observation,
although the effect is almost negligible for \my\ and \ov.

As a final check on our instrumental stability, we observed the same
probe star, BD +05$^\circ$2618, on all three occasions that we obtained
spectropolarimetry of \ov.  The variation in each Stokes parameter was
$<0.04\%$, consistent with similar experiments we have performed in
the past with repeated observations of bright stars on different
nights.  This sets an upper limit to any systematic offset between the
Lick and Keck polarimeters as well as to any night-to-night
instrumental instability.  We note that \citet{doug04dj} also found
similar results using the Kast polarimeter.  Therefore, we adopt a
value of 0.04\% for any systematic error contribution where
appropriate in the analysis below. 

\section{Results\label{IIP_results}}

As shown by \citet{li07} for \my\ and \ov, and below for \aaa, all
three objects in our study exhibit standard light-curve shapes for SNe
IIP, with the three phases of a relatively flat plateau followed by a
rapid decline phase and then a steady decline at a rate consistent
with the radioactive decay of $^{56}$Co.  This characteristic shape is 
well understood theoretically \citep{eastman94,kw09} and defines this
subclass of SNe \citep{bar79,db85}.  The supernova shock deposits
thermal energy in the envelope which is released as the hydrogen in
the envelope recombines during the plateau phase.  When the
photosphere recedes all the way back through the hydrogen layer,
recombination ceases to be a source of energy and the SN light curve
rapidly falls until the luminosity matches the instantaneous rate of
energy input from the decay of $^{56}$Co.

Both \my\ and \ov\ were discovered several months after explosion near
the ends of their plateaus.  In order to make temporal comparisons
amongst our observations of these SNe when the dates of explosion are
so poorly known, we found it convenient to define a common time to
represent the end of the plateau.  As discussed below, the timing
relative to the end of the plateau is probably the important physical
parameter for the polarization, not the time elapsed since explosion.

\citet{hamuy03} defined the end of the plateau phase, \tp, as the time
when the SN magnitude falls halfway between its value on the
plateau and the value at the start of the radioactive decay tail.  The
plateau magnitude can be difficult to uniquely define in some cases
due to the varied shapes of SN IIP light curves.  The compilation of
SN IIP photometry by \citet{dovi09} shows that some objects have
``plateau'' phases that actually rise for many days before falling
while others steadily, if slowly, decline from the beginning.  The
photometric band used to define the light curve can also have an
effect as the shapes of SN IIP light curves are a function of
wavelength.  \citet{hamuy03} used the $V$ band, but we have only
unfiltered photometry for our three objects.

Despite these uncertainties in the definition of the plateau
magnitude, \tp\ occurs during the rapid decline phase and 
hence the uncertainty in \tp\ is at worst a few days, which is
negligible for our purposes.  We measure
dates for \tp\ of 2006 Dec. 15.0, 2006 Dec. 25.5, and 2007 May 7.3 for
SNe 2006my, 2006ov, and 2007aa, respectively, based on the KAIT
unfiltered photometry.  Below, we measure \tp\ for SN 1999em in the
$R$ band (which most closely approximates our KAIT unfiltered
photometry), and our value for \tp\ differs by only 0.2 days 
from the value derived by \citet{hamuy03} in $V$, so we can conclude
that bandpass effects have negligible effect on our measured values
for \tp.  Hereafter, dates listed as ``day X'' refer to 
epochs of observation X days after the end of the plateau as defined
by \tp.

\begin{figure}
\plotone{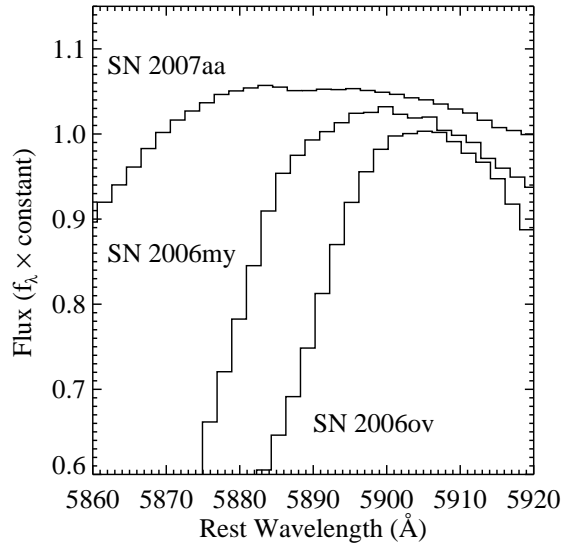}
\caption[Na I region for the three SNe]{The spectral region
  near the \ion{Na}{1} $\lambda\lambda$5890, 5896 doublet for each of
  the three SNe.  The plotted spectra are from days 38, 0, and $-50$
  for 
  SNe 2006my, 2006ov, and 2007aa, respectively.  \aaa\ shows a tiny
  dip near 5888~\AA\ that may be \ion{Na}{1} absorption in its host
  galaxy.  The other two SNe show even less evidence for host-galaxy
  absorption, indicating that we should expect negligible ISP from the
  host galaxies of these objects.
}
\label{naifig}
\end{figure}

\citet{li07} found that \my\ and \ov\ were not highly reddened and had
no detectable \ion{Na}{1}~D absorption lines from the interstellar
media of their hosts.  We confirm the lack of narrow \ion{Na}{1}
absorption in 
our spectra.  Our highest signal-to-noise ratio (S/N) spectra of each
object are plotted in Figure~\ref{naifig}.  \aaa\ does have a weak dip
near 5888~\AA\ that may be from absorption in NGC 4030.  The strength
of the feature in \aaa\ is difficult to quantify at our low spectral
resolution, especially with the curvature of the spectrum making the
continuum hard to define.  We estimate the equivalent width of the dip
to be $\sim0.08$~\AA, with an uncertainty of 
about 50\%.  If two-thirds of the absorption is due to the D1 component
of the doublet (reflecting the ratio of the oscillator strengths, as
expected in the low optical depth limit), the
relationship of \citet{munz} predicts a host-galaxy 
reddening of \ebv\ $\approx$ 0.014~mag, which corresponds to an ISP of $P
\approx 0.04$\% for typical lines of sight in the Galaxy, up to an
empirical maximum of $P \approx 0.13$\%
\citep{ser75}. SNe 2006my and 2006ov appear to have even less
absorption from their host galaxies.  Even if our estimates of the
reddening are off by a factor of a few, the conclusion that we
should expect our objects to have low host-galaxy ISP will be unchanged.

\subsection{\ov}

\begin{figure*}
\plotone{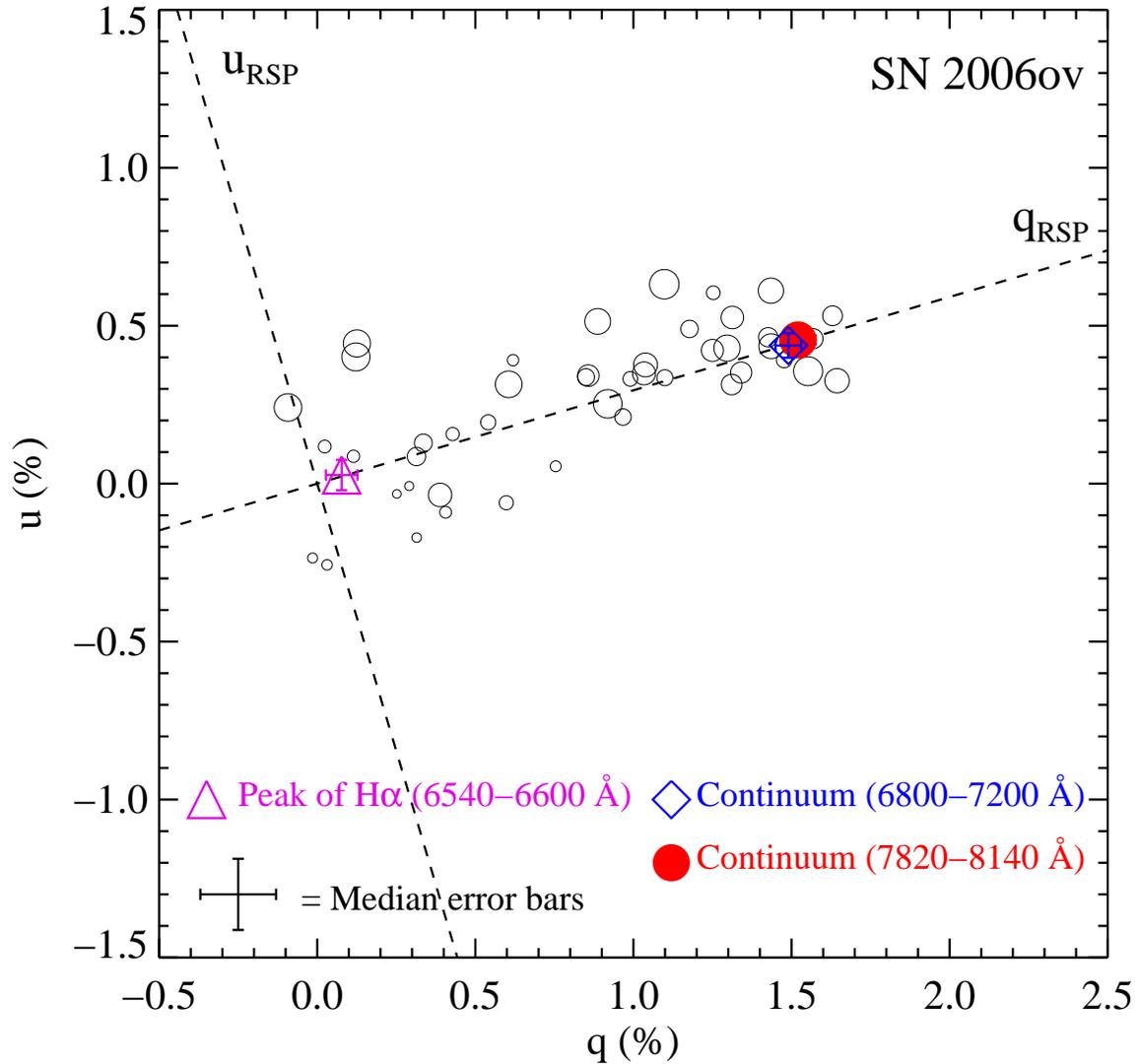}
\caption[\ov\ data in the $q-u$ plane.]{\ov\ data from 2006 Dec. 25
  (day 0) plotted in the $q-u$ plane.  The black circles represent
  100~\AA\ bins of data at wavelengths greater than 4500~\AA.  The radii
  of the circles are proportional to wavelength.  The blue diamond and
  solid red circle represent the polarization in two continuum bins
  while the purple triangle marks the peak of the \hal\ emission
  line.  The dashed lines represent the preferred, rotated set of
  Stokes parameters \qrsp\ and \ursp, chosen to align \qrsp\ with the
  two continuum bins.  The cross at ($-0.25$,$-1.3$) shows the size of
  the median error bars on the data points.
}
\label{ov_qufig}
\end{figure*}

We obtained three epochs of spectropolarimetry on \ov, two during the
steep fall off the photometric plateau and one a month later.  Our
highest S/N data were obtained from Keck on
day 0 and were fortuitously exactly synchronized with the end of the
plateau.  These data are plotted in the $q-u$ plane in
Figure~\ref{ov_qufig}, with each black circle representing a
100~\AA\ bin of data.  As described above, the S/N 
degrades to the blue, so we only plot data at wavelengths greater than
4500~\AA.  The cloud of points is clearly elongated in the positive
$q$, positive $u$ direction by an amount in excess of the error bars,
indicating substantial intrinsic supernova polarization.

We selected two broad wavelength intervals, 6800--7200 and
7820--8140~\AA, as representative of the line-free continuum (the
exact choice is justified below) and plotted the integrated
polarizations\footnote{All integrated polarizations in this paper are
calculated from the flux-weighted averages of $q$ and $u$ over the
indicated wavelength intervals.} as the blue diamond and solid red
circle in Figure~\ref{ov_qufig}.  The two continuum points are very
consistent with each other and have an average polarization (weighted
by the inverses of their variances) of $P = 1.56 \pm 0.03 \%$ at 
P.A.$=8 \fdg2 \pm 0 \fdg6$.  Many of the 100~\AA\ bins of data are
clustered around these continuum points and along the line connecting
them to the origin.  The P.A. of $8\fdg2$ therefore is significant in
this object and probably represents the projection of the SN symmetry
axis onto the plane of the sky.  Since the Stokes parameters ($q$,$u$)
are referenced to the arbitrary direction of North on the sky, it is
useful to rotate our coordinate system to align one Stokes parameter
with the symmetry axis of the SN.  We call these new coordinates
rotated Stokes parameters (RSP; Trammell et al. 1993a; Tran 1995b).
If the SN were axisymmetric, \qrsp\ would be an estimator of the total
polarization and \ursp\ would show any deviations from axisymmetry.

\begin{figure*}
\plotone{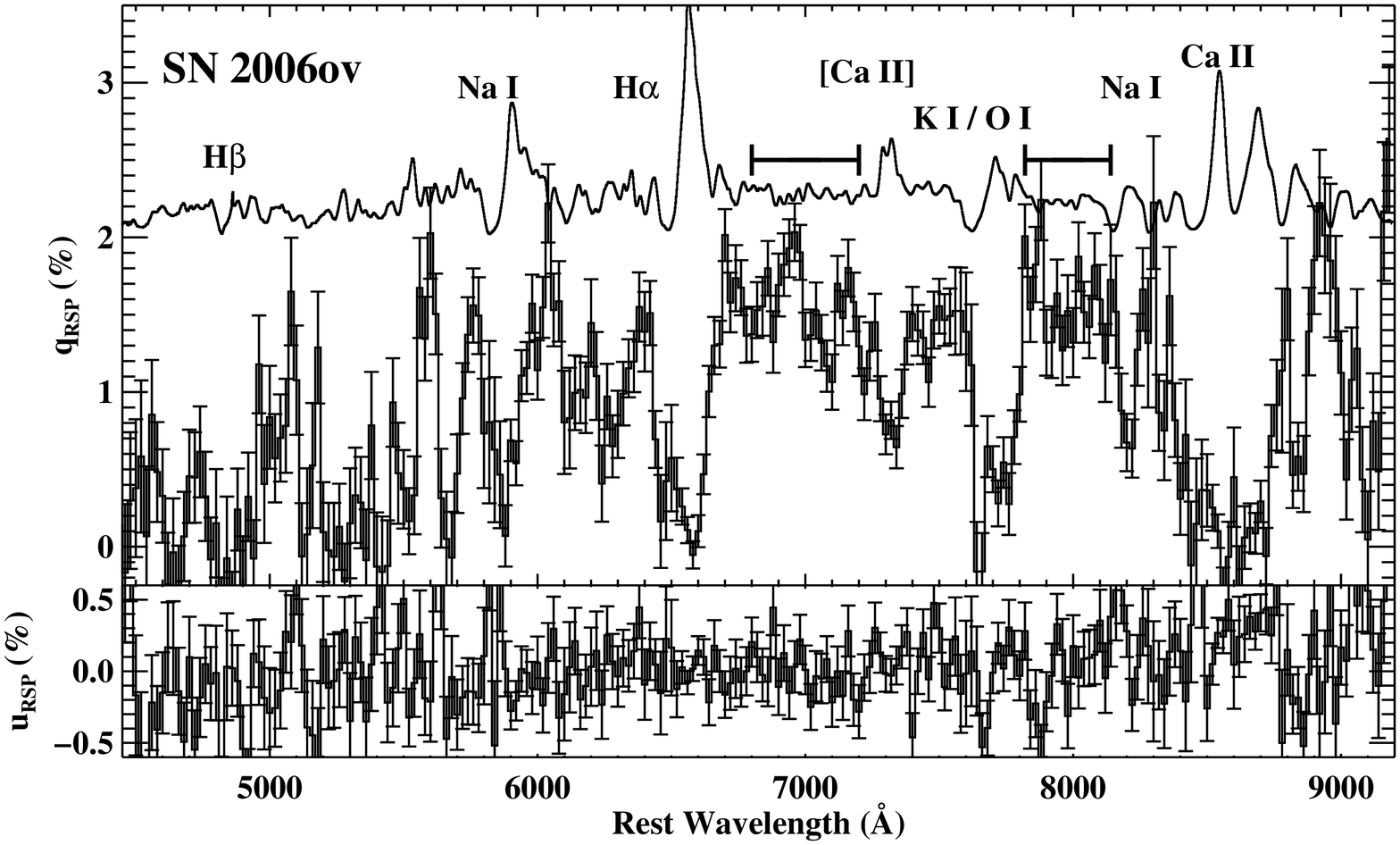}
\caption[Spectropolarimetry of \ov.]{Spectropolarimetry of \ov\ on day
  0, presented as 20~\AA\ bins in the rotated \qrsp--\ursp\ coordinate
  system so that \qrsp\ is an estimator of the polarization and
  \ursp\ is near zero.  The total-flux spectrum
  of \ov\ is overpolotted in the top panel to guide the eye, with
  several major lines labeled.  The continuum polarization between
  strong line features is very high ($\sim1.5\%$), but the
  lines result in depolarization.  The two horizontal black bars
  mark the extent of our two relatively line-free regions
  (6800--7200 and 7820--8140~\AA), which we take to be representative
  of the continuum.
}
\label{ovpolfig}
\end{figure*}

The polarization data after rotation are plotted in
Figure~\ref{ovpolfig} in 20~\AA\ bins.  The most striking aspect of
Figure~\ref{ovpolfig} is the high level of polarization ($P \approx
1.5\%$) in the continuum, with many strong, sharp depolarizations down
to almost zero in \qrsp\ present at the lines.  We integrated the
polarimetry over the wavelength range 6540--6600~\AA\ to determine
the polarization at the peak of \hal\ and measured (\qrsp, \ursp) of
(0.08\%, 0.00\%), with error bars of 0.05\% in each Stokes parameter.
The similar depolarizations present for many other lines are
convincing evidence that the host-galaxy contribution to the ISP must
be very small ($\lesssim 0.1\%$), as might be expected from the low
reddening to \ov.  To estimate the ISP, many past workers have made
the assumption that the intrinsic polarization in supernovae at the
peaks of strong lines is zero \citep{thw93, ho96, tran97}, 
but that assumption is clearly wrong in some cases \citep{maund01ig},
especially at early times when \ion{He}{1} $\lambda$6678 is blended
with \hal.  Here, the continuum polarization we measure must be almost
entirely intrinsic to \ov\ and the lines are in fact depolarized.

With that in mind, we can now justify our choice of continuum
intervals plotted in Figure~\ref{ov_qufig} and marked on
Figure~\ref{ovpolfig}.  Past work has noted the 
utility of measuring the continuum polarization in the red portion of
the optical where few line features interfere with the measurement.
\citet{doug04dj} used a wide continuum region of 6800--8200~\AA\ in
their study of SN~2004dj.  Integrating the observed polarization over
a continuum region that wide for \ov\ will clearly underestimate the
true polarization as it will include a strong depolarization present
at the \ion{K}{1}/\ion{O}{1} blend as well as a weaker one due to
[\ion{Ca}{2}] emission.  We formally get $P=1.26 \pm 0.02\%$ if we use
the wide definition for the continuum, significantly less than the
value of 1.56\% quoted above.  Therefore, we define two continuum
regions to exclude the strong lines.  The first one, 6800--7200~\AA,
is designed to avoid the red wing of \hal\ at its blue end and also 
exclude the [\ion{Ca}{2}] emission at the red end.  The second region,
$7820-8140$~\AA, is bracketed by \ion{O}{1} $\lambda$7774
and \ion{Na}{1} $\lambda\lambda$8183, 8194.

\begin{figure}
\plotone{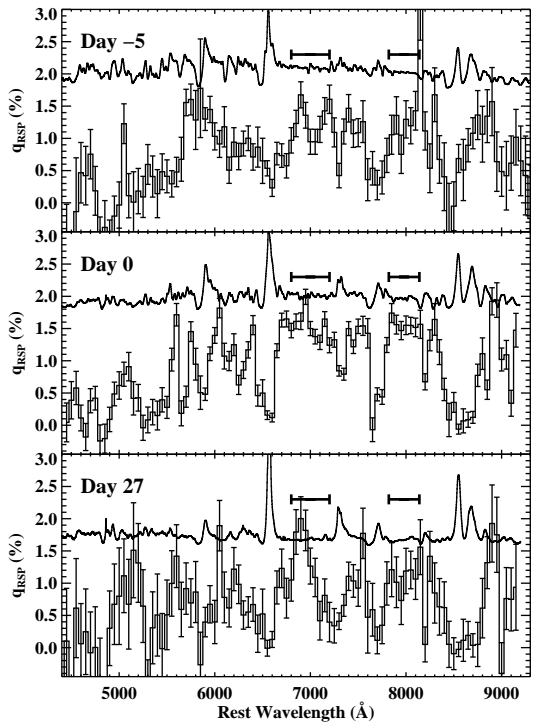}
\caption[Three epochs of polarization data for SN 2006ov.]{Three
  epochs of polarization data for SN 2006ov.  The data 
  points with error bars represent the Stokes
  parameter $q_{\rm{RSP}}$, which is an estimator of the total polarization.
  As a reference, the solid line in each figure is the total-flux
  spectrum from each epoch.  Note the high continuum polarization,
  with depressions at the location of the strong lines.  The high
  point near 8150~\AA\ in the day $-5$ polarization is spurious and due
  to relatively poor night-sky subtraction. The two horizontal black
  bars mark our two continuum regions.
}
\label{ov_compfig}
\end{figure}

We also obtained \ov\ data on two additional epochs, days
$-5$ and 27.  We have plotted a comparison of \qrsp\ on all three epochs
in Figure~\ref{ov_compfig}.  Due to the lower S/N of the additional
data, we have rebinned to 50~\AA\ per pixel.  Despite the
significant noise, the overall pattern is the same: large continuum
polarization with dips down to nearly zero present at the wavelengths of
strong line features.  The Lick data were taken only 5~days before the
day 0 Keck data and do show some significant differences,
indicating rapid evolution of the spectrum as the SN made the
transition off the photometric plateau.  For example, the
depolarizations in the wavelength range $6000-6500$~\AA\ are
more blended together at day $-5$ than at day 0.

\begin{figure}
\plotone{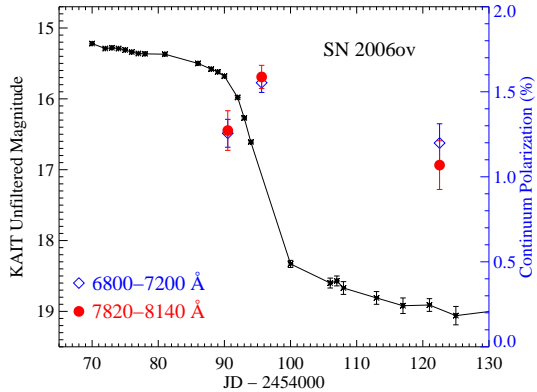}
\caption[Continuum polarization evolution in \ov.]{Continuum
  polarization evolution in \ov.  The black stars are the KAIT
  unfiltered light
  curve of \ov, with the solid line connecting the points to guide the
  eye.  The blue diamonds and solid red circles represent the
  polarization in our two continuum windows.  The end of the plateau,
  \tp, corresponds to JD $=2454095$.  The plotted error bars on the
  polarizations include a 0.04\% systematic error contribution.  The
  two horizontal black bars mark our two continuum
  regions.
}
\label{ov_lcfig}
\end{figure}

We calculated the polarization in both of our continuum regions for
each of our three epochs.  The results are plotted in
Figure~\ref{ov_lcfig} and compared to the KAIT unfiltered light
curve.  The two continuum windows gave very consistent results,
showing a significant increase in the weighted-average polarization in
only five days, from 1.25$ \pm 0.06$\% on day $-5$ to 1.56$ \pm
0.03$\% on day 0, falling to 1.15$ \pm 0.08$\% on day 27.  SN~2004dj
showed large variations in the polarization angle at the end of the
plateau \citep{doug04dj}, but the \ov\ continuum observations are
consistent with a single P.A. to within 1$\degr$.  Unfortunately,
\ov\ exploded while in solar conjunction and we were unable to obtain
observations in the month between discovery and our day $-5$ data, so
it was impossible to establish whether the high polarization
apparent at late times represented a significant change from early
times. 

\subsection{\my}

\begin{figure}
\plotone{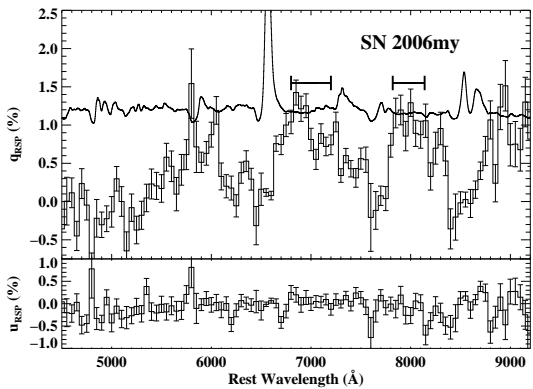}
\caption[Spectropolarimetry of SN 2006my.]{Spectropolarimetry of SN
  2006my on day 38, plotted as 50~\AA\ bins.  The Stokes parameters
  have been 
  rotated by $174\fdg1$, as described in the text.  The solid line in
  the top panel is the total-flux spectrum.  Note the high 
  continuum polarization, near 1\%, at continuum wavelengths near
  7000 and 8000 \AA.  The two horizontal black bars
  mark our two continuum regions.
}
\label{mypolfig}
\end{figure}

We obtained a single epoch of spectropolarimetry of \my\ on day 38,
2007 Jan. 21.  We analyzed the data in a similar manner to
\ov\ above.  First, we measured the weighted average polarization in
our two continuum windows to be $P=0.97 \pm 0.04\%$ at P.A. =
$174\fdg1 \pm 1\fdg2$.  Then we rotated our Stokes parameters by
$174\fdg1$ to align \qrsp\ with this polarization angle.  The results
are plotted in Figure~\ref{mypolfig}.

The data are of lower quality than for \ov\ but show broadly similar
characteristics.  High continuum polarization is seen in \qrsp, with
deep depolarizations present at the wavelengths of strong line
features.  \citet{li07} found that the extinction to \my\ was likely
to be minimal, and hence low ISP should be expected.  We determined
the polarization at the peak of \hal\ using the same window as for
\ov\ above and found (\qrsp,\ursp) of (0.03\%, 0.03\%) with an
uncertainty of 0.04\% in each Stokes parameter. 
Just as for \ov, the data for \my\ are consistent with negligible ISP
in the host galaxy.  In this object, the continuum polarization is
distinctly lower than in \ov, although it is unclear how much of the
difference is due to the later epoch of the \my\ observations.

\subsection{\aaa}

Unlike the other two objects, \aaa\ was discovered shortly after
explosion, allowing us the opportunity to obtain data at early times
and have a comparison for the late-time data.

Our first epoch of spectropolarimetry was obtained on day $-50$, well 
before the end of the plateau, and only 28 days after discovery.  As
discussed above, \citet{fol07} described a spectrum taken one day
after discovery as being similar to that of SN~1999em at 20 days after
explosion, implying an epoch for our first observation of $\sim47$
days after explosion.  We used the SuperNova IDentification code of
\cite{snidref} to cross-correlate our day $-50$ spectrum with a
library of supernova templates, holding the redshift fixed.  The top
three individual SNe with spectral matches (ignoring spectra on
multiple dates per object) were SNe 2004et, 2004dj, and 2006bp at 56,
56, and 58 days (respectively) after their estimated explosion dates
\citep{sahu04et,vinko06,quimby06bp}. 

These estimates are all consistent with \aaa\ having a normal plateau
duration of $\sim100$ days, with our first set of data being taken
near the middle.  Our second epoch, at day $-22$, was taken late in
the plateau stage, but before the beginning of the steep decline
phase, while our third epoch, on day 3, was taken during the steep
decline and just before the beginning of the radioactive-decay tail.

\begin{figure}
\plotone{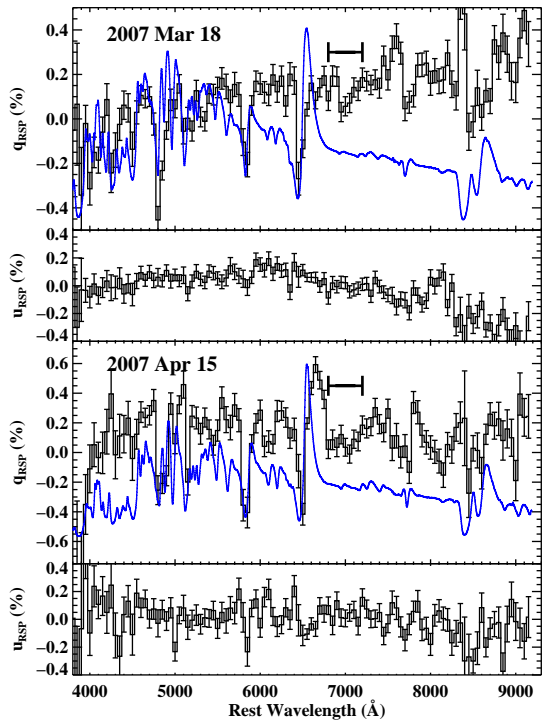}
\caption[Early-time spectropolarimetry of \aaa.]{Early-time
  spectropolarimetry of \aaa\ obtained while the object was on the
  plateau, on days $-50$ and $-22$.  The polarization data have been
  rotated by 172$\fdg$4 to align \qrsp\ with the direction of the
  late-time continuum polarization.  The blue solid lines in the
  background of each \qrsp\ plot are the total-flux 
  spectra from each date.  Note the weak continuum polarization with
  significant line polarization features.  The ``droop'' in \ursp\ to
  negative values in the 2007 March 18 polarization data at
  wavelengths greater than 8200~\AA\ is unexplained and may not be
  real, although none of the standard or probe stars observed that
  night showed such an effect. In any case, none of the results 
  presented in this paper rely on data from that portion of the spectrum. 
  The horizontal black bar in each \qrsp\ panel
  marks our continuum region.
}
\label{aa_earlyfig}
\end{figure}

\begin{figure*}
\plotone{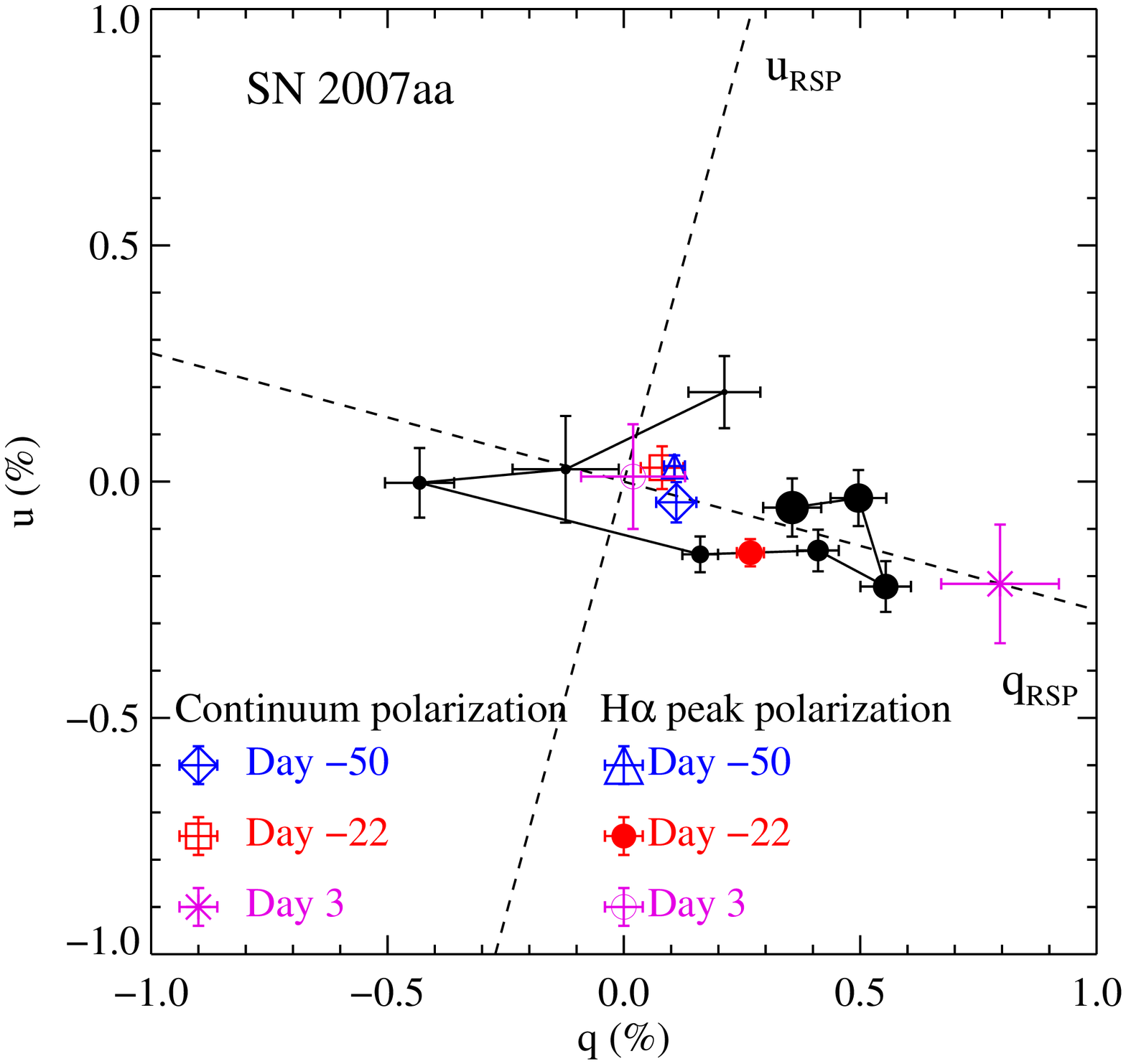}
\caption[$q-u$ plot of \aaa\ polarization.]{Polarization data for
  \aaa\ plotted in the $q-u$ plane.  The solid black circles connected
  by a line are from the day $-22$ data and are 50~\AA\ bins in
  wavelength from 6400 to 6750~\AA, representing a velocity range of
  $-7500$ to 8400~\kms\ relative to \hal.  The points form a loop in the
  $q-u$ plane.  The radius of each circle is proportional to
  wavelength, so moving across the line profile from blue to red
  corresponds to moving around the loop in a counterclockwise
  direction.  Also plotted are continuum polarization points 
  (integrations over 6800--7200~\AA) and a set of points
  representing the peak of the \hal\ emission line (integrations over
  6540--6600~\AA).  Note the low polarizations in the \hal\ peaks and
  in the continuum for the first two epochs, which together imply
  minimal ISP in this object.  The day~3 continuum
  polarization represents a significant jump in polarization.  The
  dashed lines represent the rotated \qrsp$-$\ursp\ coordinate system,
  designed to have the \qrsp\ axis go through the day~3 continuum
  point.
}
\label{aa_qufig}
\end{figure*}

The two plateau epochs of spectropolarimetry are plotted versus
wavelength in Figure~\ref{aa_earlyfig} and in the $q-u$ plane in
Figure~\ref{aa_qufig}.  We have applied a rotation of 172$\arcdeg$ to 
the data in Figure~\ref{aa_earlyfig} in anticipation of the late-time
results.  A low level of 
continuum polarization at early times in \aaa\ is clear 
from Figure~\ref{aa_earlyfig}.  At both epochs, \ursp\ was near zero and
\qrsp\ ranged over $0-0.2$\% in the continuum.  Line polarization
features are clearly present at the Balmer, \ion{Na}{1}, \ion{Fe}{2},
\ion{Ca}{2}, and \ion{O}{1} lines.  The strongest line polarization
feature, at \hal, had an extent of about 0.4\% in the first epoch
and grew to $\sim1$\% in the second epoch.  Close inspection of
\ursp\ in the second-epoch data shows small variations at the
positions of the line features.  This is a hint that the line features
indicate deviations from axisymmetry.

To examine this in more detail, we have plotted the data points in the
\hal\ line from day $-22$ in Figure~\ref{aa_qufig} as the solid black
circles.  The data points are 50~\AA\ bins in wavelength from
6400 to 6750~\AA, representing a velocity range of
$-7500$ to 8400~\kms\ relative to \hal.  The radius of each circle is
proportional to wavelength, so the smaller points are in the blue wing
of the profile and the larger circles are from the red wing.  The
\hal\ points show a spread in both $q$ and $u$
in a manner inconsistent with a simple axisymmetric geometry.  In
axisymmetry, all the points would lie along a line in $q-u$ space.
Instead, the points show a $q-u$ ``loop.''  As one moves across the
\hal\ profile from red to blue, one moves in a counterclockwise
direction around the loop.

Such loops were first identified in supernova spectropolarimetry by
\citet{cropper88} in the \hal\ and \ion{Ca}{2} near-infrared (NIR)
triplet lines of SN~1987A.  The interpretation of the loops in
SN~1987A has remained obscure for over 20 years despite the high
quality of the data \citep{ww08}.  Subsequent investigations 
have found $q-u$ loops in systems as different from SN~1987A as SNe
IIn \citep{hoffman97eg}, stripped-envelope SNe
\citep{maund05bf,maund01ig}, and even SNe Ia \citep{ka03,wa03,me08}.
\citet{ka03} found that loops 
were a generic consequence of systems containing multiple components
with misaligned axes of symmetry and constructed a few examples with
very different geometries.  The relevance of these models to SN~II
atmospheres during the plateau stage is unclear as both the
photosphere and \hal-forming region are within the hydrogen envelope
of the progenitor.  \citet{ch92} has a more promising model,
specifically designed for SN~1987A, which interprets the \hal\ loops
as evidence of a small number of individual clumps of $^{56}$Ni
creating pockets of high ionization and \hal\ excitation.  At
different radial velocities, we see the effects of different clumps
causing the net polarization angle to rotate with wavelength.

We defer additional analysis of the line polarization features in
\aaa\ to future work on the polarization of SNe~II at early times
because our focus here is on the continuum polarization and its
temporal evolution.  For now, we will note the following points.
First, the long axis of the \hal\ loop on day $-22$ is mostly aligned
with the P.A. of the late-time continuum polarization, possibly
indicating a common symmetry axis.
Second, the total polarization excursion in the line is $\sim1$\%
at an epoch where the continuum polarization is likely to be very much
lower, as discussed below.

We will now discuss the continuum polarization in detail.  We found
our low-S/N late-time polarization data in the final epoch to be
unreliable near 
8100~\AA, so we have focused only on the bluer of our two continuum
polarization windows, 6800--7200~\AA.  Figure~\ref{aa_qufig} has our
continuum polarization measurements plotted for each of our three
epochs of observation.  As was clear from Figure~\ref{aa_earlyfig},
the observed continuum polarization on the plateau was very low.  We
measure formal polarizations of 0.11 $\pm$ 0.04\% and 0.08 $\pm$
0.05\% on days $-50$ and $-22$, respectively.  By contrast, the
polarization after the end of the plateau jumped to 0.83 $\pm$ 0.12\%
on day 3.

\begin{figure}
\plotone{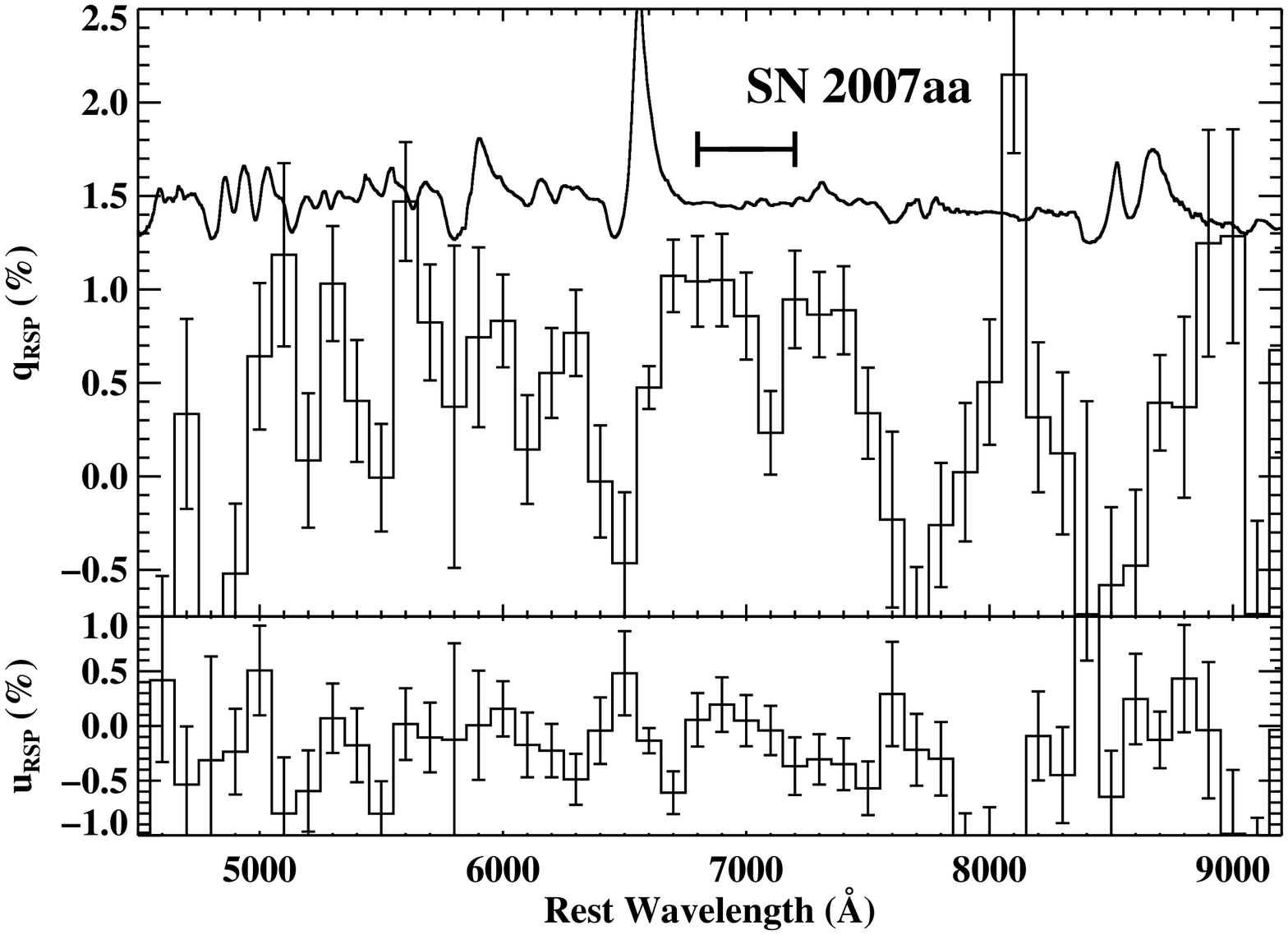}
\caption[Late-time spectropolarimetry of \aaa.]{Late-time
  spectropolarimetry of \aaa\ from day 3, presented in 100~\AA\ bins.
  The solid line in the upper part of the top panel is the total-flux
  spectrum to guide the 
  eye.  High continuum polarization near 1\% is seen in \qrsp\ with
  dips down near zero at the lines, while \ursp\ is consistent with
  zero.  The spike in \qrsp\ and dip in \ursp\ near 8100~\AA\ are
  spurious and are caused by systematically poor night-sky subtraction
  near that wavelength.  The horizontal black bar
  marks our continuum region.
}
\label{aa_latefig}
\end{figure}

The late-time spectropolarimetry data of \aaa\ from day 3 are plotted
in Figure~\ref{aa_latefig}.  The data are extremely noisy due to the
faintness of the object at this time, but when binned up to
100~\AA\ per pixel the basic appearance of the data is qualitatively 
similar to that shown previously for \ov\ and \my.  The continuum
polarization is near 1\% in \qrsp, but there are broad dips down to
near zero polarization at the locations of strong lines such as \hal,
with \ursp\ centered near zero.  The polarization we measure 
at the peak of \hal\ is (\qrsp, \ursp) = (0.02\%, 0.02\%), with 0.1\% error
bars in each Stokes parameter, consistent with zero, as shown
in Figure~\ref{aa_qufig}.

A few of the dips in polarization at the lines appear to reach
negative values of \qrsp, especially near 7650~\AA\ and 8400~\AA, but these
are only marginally statistically significant.  The wavelength bins
that are the most negative are those with the least flux from the SN
(due to absorptions from the telluric A band and the \ion{Ca}{2} NIR
triplet P-Cygni feature) and hence the lowest S/N, magnifying
any systematic uncertainties in the sky subtraction.  An integration
over $8500-8800$~\AA, representing the emission component of the
\ion{Ca}{2} NIR triplet, gives (\qrsp, \ursp) = ($-0.09\pm0.17$\%,
$-0.11\pm0.17$\%), which like the polarization at the peak of \hal\ is
completely consistent with zero.

If we interpret the late-time polarization of \aaa\ as being similar
to that of \ov\ and \my, then the contribution from ISP should be low in this
object as well, given the low polarization in the lines.  Further
evidence for this can be seen in 
Figure~\ref{aa_qufig}, where we have plotted the peaks of the
\hal\ lines for all three epochs (integrated over the same
$6540-6600$~\AA\ interval).  The \hal\ peaks in the first and last
epoch are near zero.  The \hal\ polarization in the second epoch is
higher, but still low (0.3\%).  That is also the epoch
showing clear evidence of strong polarization throughout the line, so
perhaps a non-zero polarization at the peak is to be expected,
particularly if an aspherical excitation distribution is present
\citep{ch92}.  Small amounts of ISP have a similar effect
to changing the origin in the $q-u$ plane, so there is still a
possibility of a pathological ISP canceling out some intrinsic
supernova polarization.  However, taken together with the low
continuum polarization measured on the plateau, the simplest
interpretation is that \aaa, like \ov\ and \my, suffers minimal ISP
($\lesssim0.1$\%).

\begin{figure}
\plotone{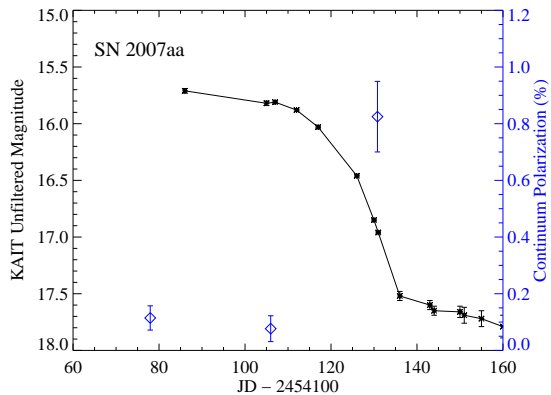}
\caption[Polarization and light curve of \aaa.]{Time evolution of the
  continuum polarization of \aaa\ compared to the KAIT unfiltered
  light curve (black stars connected by a line).  The blue diamonds
  represent the continuum polarization integrated over
  6800--7200~\AA.  The derived time of the end of the plateau, \tp,
  is JD = 2454227.8.
}
\label{aa_lcfig}
\end{figure}

We have taken the continuum polarization measurements and plotted them
versus time in Figure~\ref{aa_lcfig}, along with the KAIT unfiltered
light curve.  The polarization shows a strong increase from the
plateau stage to the late-time observations taken during the rapid
decline phase.  Even if we are mistaken about
the ISP being negligible, the observation that there was a sharp
change in the polarization associated with the end of the plateau is a
robust result.  This is strongly reminiscent of the jump in
polarization seen in SN~2004dj \citep{doug04dj}.  In \aaa,
however, there was evidence of asphericity in the form of line
polarization on the plateau, while SN~2004dj did not even show any
line polarization features before the sudden increase in continuum
polarization.

\section{Discussion\label{IIP_discussion}}

The spectropolarimetry data presented above for the three objects
after the end of the plateau are all qualitatively similar with each
other and with SN~2004dj \citep{doug04dj}.  The continuum polarization
for each object is high, but flat and constant with wavelength between
the strong line features in the red.  This is a signature of the 
wavelength-independent nature of Thomson scattering.  The lines
produce large depolarizations and many overlapping lines significantly
decrease the net polarization in the blue.  The minimal polarizations
at the peaks of \hal\ and lack of significant interstellar \ion{Na}{1}
absorptions are consistent with negligible ISP in all three objects,
and therefore the large continuum polarizations are intrinsic to the
SNe.

Classically, the expected effect of lines forming in front of an
aspherical electron-scattering photosphere is to produce ``inverse
P-Cygni'' line polarization features \citep{jef89}.  Line emission
near zero velocity produces a dip in polarization near the line center,
and selective blocking of forward-scattered light produces a
polarization peak in the blueshifted absorption trough.  Higher
polarizations can also be found in the red wing of the line as the
opaque photosphere obstructs the observer from seeing unpolarized line
emission from the far side of the ejecta.  This simple picture
provides an explanation for the line polarization features in
photospheric-phase spectropolarimetry of SN~1987A and SN~1999em
\citep{je91,je91catalog,le01}, and possibly our early data for
\aaa\ as well.  However, we do not see emission peaks blueward of 
\hal\ or the \ion{Ca}{2} NIR triplet in our highest S/N late-time data
(Fig.~\ref{ovpolfig}).  Instead, the polarization is simply low
relative to the continuum throughout the line profile.  There are a few
contributing effects.

The first is simple flux dilution. The continuum photosphere and
associated polarizing electrons are located relatively deep in the
supernova ejecta and intrinsically unpolarized line emission from the
optically thin outer layers of the ejecta travels unimpeded to the
observer and dilutes the observed polarization from 
the continuum value.  As an example, the [\ion{Ca}{2}] $\lambda$7300
doublet in the day 0 \ov\ data represents emission from the
low-density parts of the outer ejecta.  The line emission 
strengthens over time as the ejecta thin out and the supernova begins
the transition to the nebular phase.  This line has a flux peak a
little less than a factor of $\sim$2 above the continuum (the continuum 
flux level is hard to define at this epoch because of many overlapping
P-Cygni features), and the polarization dips from 
$1.3-1.4$\% in the nearby continuum pixels down to $\sim0.8$\% in the
line.

The second reason is that photons scattering in lines are generally
depolarized \citep{ho96}.  While the outer layers of the ejecta are
optically thin in the continuum, it seems likely that the optical
depth in the strong lines is still significant.  The depolarizations
over the full line profile from lines that are strong in the cool
outer ejecta, such as \ion{Na}{1}, \ion{Ca}{2}, and \hal, are likely
manifestations of this effect.  Absorption and reemission in the line
causes the photons to lose the directional information previously
imprinted upon them by electron scattering.  The net effect of a large
number of overlapping lines in the blue is to result in a decrease in
polarization at those wavelengths \citep{ho01}.  In addition, the
line-formation region is very geometrically extended compared to
the electron-scattering photosphere, so the selective blocking of
forward-scattered light is not an important effect.

A few of the strong depolarizations do not correspond to obvious
strong features in the total-flux spectrum.  An example is the
depolarization visible near 5700~\AA\ in Figure~\ref{ovpolfig}, which
was also present in SN 2004dj \citep{doug04dj}.  It is unclear which
weak spectral feature is responsible for that depolarization and why
the many other weak spectral features do not also result in similarly
strong depolarizations.

The time evolution of the polarization is of significant interest.
SNe~IIP generally do not exhibit strong polarization shortly after
explosion \citep{ww08}, and yet SN~2004dj had a strong jump in
continuum polarization near the end of its plateau \citep{doug04dj}.
The main motivation for this study was to determine whether SN~2004dj
was unique, or if other SNe~IIP exhibit similar polarization
behavior.  We have shown time-series plots for the two objects in our
study with multi-epoch polarimetry in Figures~\ref{ov_lcfig} and
\ref{aa_lcfig}, but \aaa\ was the only object with data taken on the
plateau to establish the low early-time polarization.

To make further progress, we searched for SN~IIP polarimetry in the
literature.  As discussed in \S1, there are surprisingly
few published datasets of SN~IIP spectropolarimetry in the refereed
literature.  Since we are interested in the intrinsic continuum
polarization, we must also exclude objects with large ISP corrections
because estimates of ISP are uncertain and require making
assumptions.  SN 1999em \citep{le01,ww08} and SN 2004dj 
\citep{doug04dj} are the only objects from the published literature to
meet our criteria.  Of the three objects in the study by \citet{lf01},
SNe 1997ds and 1999gi both had large and uncertain ISP contamination,
and SN~1998A was a SN~1987A analog of unclear relevance to our
SNe~IIP.  \citet{lf05} also plotted continuum polarization versus time
for several SNe~IIP, but three of the five objects in their sample had
ISP corrections of 1\% or more.

\begin{figure*}
\plotone{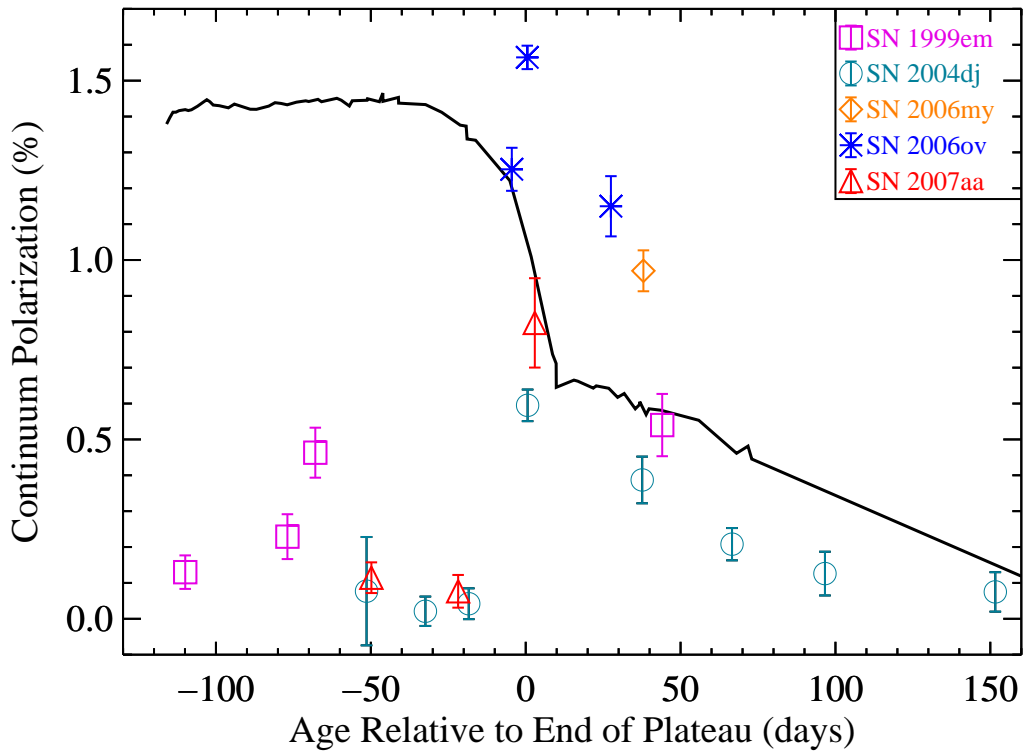}
\caption[Compilation of SN IIP continuum polarizations.]{Compilation
  of SN IIP intrinsic continuum polarization measurements after
  correction for ISP.  The SN~1999em data (purple squares) are
  from \citet{le01} and the SN~2004dj data (cyan circles) are from
  \citet{doug04dj}, reanalyzed in a manner similar to the current
  objects (see text for details).  The SNe 2006ov (blue stars), 2006my
  (orange  
  diamond), and 2007aa (red triangles) data points are from this
  work.  The abscissa represents time in days relative to the end of
  the plateau, \tp, for each object, as defined in the text.  We used
  $R$-band photometry from 
  \citet{doug04dj} and \citet{vinko06} to derive \tp\ for SN~2004dj of
  2004 Oct. 12.4, and $R$-band photometry from \citet{doug99em} and
  \citet{hamuy99em} were used to derive \tp\ for SN~1999em of 2000 Feb
  23.3.  A systematic error contribution of 0.04\% has been added in
  quadrature to all points to account for potential night-to-night
  instrumental variations, except for the SN 2004dj data, which
  already included a systematic error contribution.  The black solid
  line is 
  the well-sampled $R$-band light curve of SN~1999em to guide the eye.
  The data are consistent with a sharp increase in the continuum
  polarization in the last 10--20 days before the end of the plateau.
}
\label{allfig}
\end{figure*}

We collected photometry for the two objects from the literature and
determined the date of the end of the plateau for each.  The combined
set of polarization versus time since \tp\ is shown in
Figure~\ref{allfig}.  We must caution the reader that the data shown
in this Figure are necessarily very heterogeneous.

We also reanalyzed the published polarization data in order to make
our comparisons as uniform as possible.
The published SN~1999em data generally do not extend sufficiently far
to the red to include our favored continuum regions.  We calculated
the continuum polarization using as much of our 6800--7200~\AA\ window
as was available in each observation from the dataset of \citet{le01}.
The 1999 Nov. 5 and 2000 Apr. 5/9 data included the full wavelength
range, while for the 1999 Dec. 8 and 17 data we integrated from
6800~\AA\ to the red wavelength end of the data.  The 1999 Dec. 17
continuum polarization is the only one of the four values to
differ from the synthesized $V$-band polarizations quoted by
\citet{le01} by more than the error bars, possibly due to the limited
wavelength range (the data from that night only extend redward to
6850~\AA).  The data have not been corrected for ISP because the
effect is believed to be small ($\sim0.1$\%; Leonard et al. 2001).  

As discussed above, applying the very wide continuum region used by
\citet{doug04dj} in their study of SN~2004dj to our \ov\ data would
reduce the overall polarization measured on day 0 from 1.56\% to
1.26\%. 
To correct for this effect, we obtained the SN~2004dj data from
D. Leonard and calculated the polarization in our two continuum
windows in an identical manner to the other objects in this paper.
Therefore, the peak continuum polarization plotted in
Figure~\ref{allfig} is now 0.60\%, $\sim$10\% higher than the
published value of 0.56\% given by \citet{doug04dj}.

Despite these caveats, we believe the basic trend shown in
Figure~\ref{allfig} is sound.  At early times, SNe~IIP are seen to
exhibit low polarization \citep{ww08}.  The large, extended,
slowly rotating hydrogen envelopes of their progenitors are unlikely
to be substantially aspherical prior to explosion.  Even an aspherical
initial explosion will become more spherical when propagating outward
through the remainder of the star \citep{cs89,kifonidis06}.  At the
end of the plateau, the thick hydrogen envelope becomes mostly
optically thin, allowing us to see deep into the ejecta and observe
the aspherical core of the explosion.  In addition, the observed
polarization for a given geometry is expected to be maximized when the
optical depth to electron scattering is nearly unity \citepeg{ho91}.
To be fair, the dramatic jump in polarization was only directly seen
in SN~2004dj and \aaa.  However, \ov\ does show a consistent increase
in polarization of 0.3\% in both of our continuum regions over the
course of just five days, indicating very rapid changes in the
continuum polarization. 

It is instructive to compare to the supernova with the most densely
sampled polarimetric data, SN~1987A.  \citet{je91catalog} presented
ISP-corrected broad-band polarimetry from a number of sources
\citep{bar88,mendez88}.  The $BVRI$ polarizations do show a common
trend of rising polarization at late times, with a possible spike in
the $B$ and $V$ polarization percentages near the beginning of the
radioactive-decay tail.  While the SNe~IIP in Figure~\ref{allfig}
show declining polarization after the start of the radioactive tail,
the SN~1987A data continue rising up to $\sim1$\% at 200--300 days
after explosion.  The Type IIb SN~2001ig \citep{maund01ig} also showed
a distinct jump in polarization associated with the hydrogen layer
becoming optically thin and the photosphere receding into the He
core.

Prior to the publication of the SN~2004dj polarization results,
\citet{ch05} noticed unusual line profiles present in that object in
the nebular phase.  The \hal, \hb, [\ion{O}{1}], and [\ion{Ca}{2}]
lines all exhibited a double-peaked structure with the main peak
initially blueshifted by about 1600~\kms\ and moving redward with
time, along with a similar peak on the red side of the profile.
\citet{ch05} modeled these line profiles in terms of a bipolar 
$^{56}$Ni component protruding from a more spherical ``cocoon'' that
produced an ionization and excitation asymmetry.  \citet{ch06} found
that minimal adjustments to the initial model could roughly match the
polarization results at the second nebular epoch, although his models
did not predict the decline in polarization after the beginning of the
radioactive-decay tail seen in the data or the different polarization
angle seen in the first nebular dataset relative to the later data.

\begin{figure}
\plotone{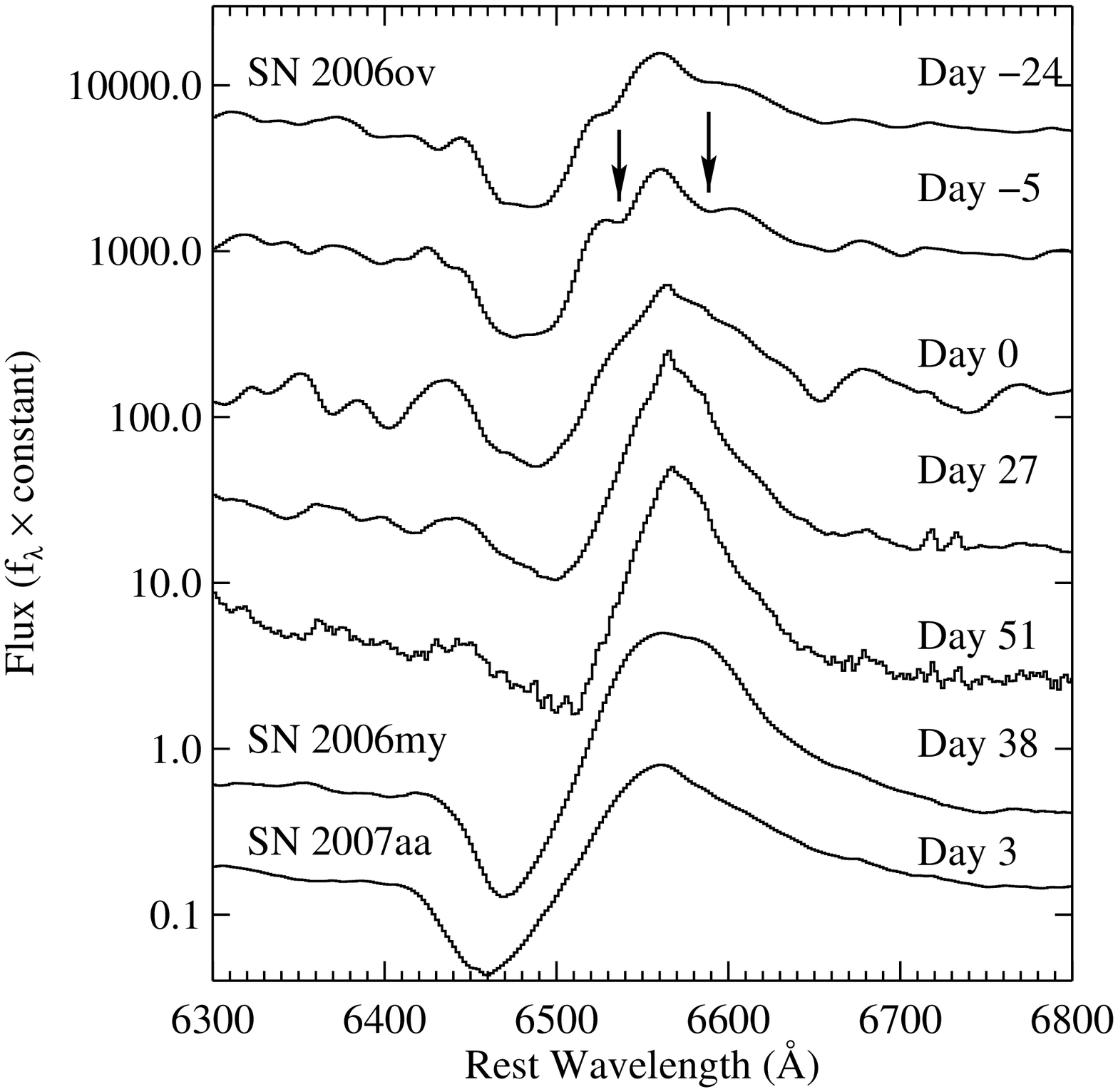}
\caption[Late-time \hal\ line profiles.]{Late-time \hal\ line
  profiles.  The spectra are labeled by their ages relative to \tp.
  The bottom two spectra are of SN~2006my and 2007aa, while the
  remainder are of \ov.  The day $-24$ spectrum of \ov\ is from
  \citet{li07}, while the day 51 spectrum was acquired on 2007
  Feb. 14.655 with LRIS using a setup similar to that for the other
  spectra. 
  Two notches (marked with arrows) are seen in the earliest
  \ov\ spectra near 6535 and 6590~\AA.  At later times, none of the
  objects shows significant asymmetries in the \hal\ line profile.
}
\label{halfig}
\end{figure}

While asphericities in the $^{56}$Ni distribution are likely
related to the polarization we observe in the three objects in
our sample, none of them exhibits late-time line profiles similar to
those of
SN~2004dj.  The late-time \hal\ profiles of our objects are shown in
Figure~\ref{halfig}.  SNe 2006my and 2007aa have mostly rounded
\hal\ profiles, with \my\ showing a small notch or dip near 6575~\AA.
The early-time spectra of \ov\ do show a pair of double notches, near
6535~\AA\ and 6590~\AA\ in the day $-5$ spectrum, corresponding to
velocities of $-1250$ and 1200 \kms\ with respect to \hal.  The
notches disappeared in just five days and the line profiles after the
end of the plateau were normal, by contrast with SN~2004dj, 
whose \hal\ line-profile asymmetries remained prominent for at least
330 days after explosion \citep{ch05}.  The blue and red notches in
the \hal\ profile of \ov\ before the beginning of the
radioactive-decay tail are reminiscent of the so-called ``Bochum
event'' in SN~1987A \citep{bochum}, although at significantly lower
velocities ($\sim1200$~\kms\ versus $\sim4000$~\kms\ in SN~1987A).
Similar notches have been seen in other SNe~IIP near the end of the
plateau (e.g., at $v\approx2500$~\kms\ in SN 1999em; Leonard et
al. 2002a) and do likely indicate asphericities or clumps in the
$^{56}$Ni distribution \citep{ch92,utrobin95}. 

Given the evidence for a potentially clumpy $^{56}$Ni distribution, it
is worth considering whether clumps of $^{56}$Ni or other deformations
of the core-envelope boundary could be responsible for the continuum
polarization that we measure, rather than an aspherical photosphere
in the core.  \citet{doug04dj} hypothesized that the
change in the polarization angle seen from the first to the second
epoch after the end of the plateau in SN~2004dj was due to some
non-axisymmetric portion of the $^{56}$Ni distribution beginning to be
uncovered before the whole 
core as the photosphere receded in the ejecta.  \citet{ch06} proposed
an alternative explanation, which he termed a ``spotty photosphere,''
invoking an inhomogeneous distribution of the elements at the
hydrogen-helium boundary due to fluid instabilities at that interface
during the explosion \citep{muller91}.  These inhomogeneities would
result in brightness and opacity variations that would be manifested
as a nonzero net polarization that suddenly appears when the
overlying layers become optically thin and the photosphere passes
through the core-envelope boundary, even if the core of the SN were
essentially spherical on large scales.

Our data for these three objects do not show any angle variations of
the type seen in SN~2004dj.  The polarization angles for \ov\ are
identical for all three epochs from day $-5$ through day 27.  While
the continuum polarization for \aaa\ at early times is sufficiently
small that the polarization angle is hard to reliably measure (a small
amount of ISP could result in dramatic changes in the inferred angle),
the long axis of the $q-u$ loop seen in the \hal\ line on day $-22$
(Fig.~\ref{aa_qufig}) is approximately aligned with the late-time
continuum polarization.  The combination of low continuum polarization
and significant line polarization (as seen in day $-22$ in \aaa) may
indicate that the \hal\ excitation is more sensitive to the underlying
aspherical $^{56}$Ni distribution than the
continuum and hence the alignment with the late-time polarization is
not accidental.  In addition, the continuum polarization in 
SN~1999em remained at a constant P.A. over the course of the first
161 days after explosion \citep{le01}.  The polarization angle
variation of SN~2004dj appears to be an outlier compared to these
other objects, or else the first observation after the end of the
plateau was very fortunately timed to catch the polarization angle
while it was rapidly varying.

The simplest explanation for the constancy of the polarization angles
is that these SNe~IIP have highly aspherical cores surrounded by
hydrogen envelopes that are more spherical.  Thus, we
conclude that our late-time continuum polarization measurements are
most easily understood as good tracers of the underlying asphericity
of the photosphere in the core, and not of a ``spotty photosphere'' or
other boundary effect that only produces a transient polarization
signal when the photosphere enters the core.   Observations taken
well before the photosphere recedes through the boundary between the
hydrogen-rich envelope and the helium-rich core already show signs of
the asphericity that is present at late times.

Our maximum continuum polarization of 1.56\% in \ov\ on day 0 implies
a minimum axis ratio of 1.45:1 in the context of the aspherical
electron-scattering atmospheres of \citet{ho91}, if the object were
viewed in the equatorial plane, and possibly higher if the symmetry
axis were less inclined to the line of sight.  Despite the lower
continuum polarization observed in SN~2004dj, \citet{doug04dj} derived
a similar axis ratio for the core of that object after accounting for
the preferred inclination of \citet{ch05}.  

\begin{figure}
\plotone{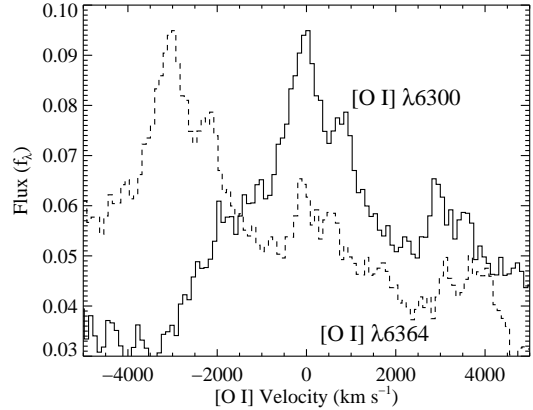}
\caption[Late-time oxygen line profiles in SN 2006ov.]{Late-time
  [\ion{O}{1}] line profiles in \ov\ from day 51.  The solid line
  shows velocities relative to $\lambda$6300 and the dashed line shows
  velocities relative to the $\lambda$6364 component of the doublet.
  Two peaks are observed in both, near 0 and 850 \kms\ 
  for $\lambda$6300.  These are likely a sign of large-scale clumps in
  the oxygen-rich inner core, perhaps as a result of
  Rayleigh-Taylor instabilities in the explosion.
}
\label{oifig}
\end{figure}

\ov\ also shows some evidence for inner asphericity in the late-time
[\ion{O}{1}] line profiles.  The profiles of the $\lambda\lambda$6300,
6364 doublet in our latest spectrum (day 51) are shown in
Figure~\ref{oifig}.  Although the spectrum has not yet become fully
nebular and the [\ion{O}{1}] doublet does not yet completely dominate
its portion of the spectrum, a double-peaked structure is apparent.
The peaks in $\lambda$6300 are at velocities near 0 and 850 \kms.
Confirmation that this represents the oxygen distribution and not
contamination by some other line can be seen from the profile of
$\lambda$6364, which shows a peak near zero velocity as well as one to
the red; however, the redder peak in $\lambda$6364 is closer to 500
\kms.

Double-peaked [\ion{O}{1}] profiles have been seen in
stripped-envelope supernovae and attributed to a toroidal oxygen
distribution \citep{maz03jd,maeda08,modjaz08}, but the
velocity splittings of the peaks in those objects are significantly
higher.  Alternative geometries are also possible
\citep{taub09,mili09}.  Here, a more likely  
explanation is that we are seeing large-scale clumps or blobs in the
inner oxygen distribution, especially given that the main peak is at
zero velocity.  Such oxygen clumps have been seen in stripped-envelope
supernovae \citep{fs89,ma00,silv09}, SN~1987A \citep{stathakis91}, and
the SN IIP 2002hh \citep{pozzo06}.  The clumps are usually attributed
to Rayleigh-Taylor instabilities created at the interface of the
oxygen-rich and helium-rich layers in the core during the explosion
\citep{muller91}, although other large-scale fluid motions during the
explosion are possible.

\section{Conclusions}

We have presented late-time spectropolarimetry of three SNe~IIP.  All
three objects had large continuum polarizations after the end of the
photometric plateau ($0.8\lesssim P \lesssim 1.6$\%), while \aaa\ had
contrastingly smaller continuum 
polarization on the plateau ($P \approx 0.1$\%).  These results
confirm that the polarization evolution of SN~2004dj \citep{doug04dj}
was not unique and that highly aspherical cores exist in
SNe~IIP.  This follows the trend identified by \citet{wa01}
that the degree of polarization in core-collapse SNe generally
increases with time.

It is important to note that an intrinsic continuum polarization of
1.56\% is quite high\footnote{We caution that the precise polarization
  value is dependent on the definition of the continuum regions.}.
Although 
there are extreme cases such as the SN Ic 1997X \citep{wa01}, very few
core-collapse SNe exhibit intrinsic continuum 
polarizations higher than this.  Most stripped-envelope SNe have
significantly lower continuum polarization (after correction for ISP),
including even extreme examples such as the well-studied broad-lined
SN~Ic 2002ap \citep{kawabata02ap,le02,wang02ap}.  The line
polarization features in some objects are stronger, however.  SNe~IIP
are the single most common form of core-collapse supernova and
represent the explosions of the most common massive stars, those at
the low-mass end of the range of supernova progenitors.  Our
observations of these objects demonstrate that large degrees of
asphericity are generic ingredients in even ordinary core-collapse
explosions and are not just limited to various unusual or extreme
objects.

This observational view parallels recent developments in the
theoretical modeling of core-collapse SNe.  Modern simulations of
supernova explosions have revealed a rich phenomenology of 
multi-dimensional effects \citep{sasiref,burrows06}.  Extreme models
invoking jets \citep{khokhlov99} have had some success explaining the
polarization evolution and other tracers of asphericity in
core-collapse SNe \citep{hkw01,couch09}, although the mechanism by
which jets would be launched in ordinary SNe IIP remains obscure, as
jet formation is usually thought to require atypically high angular
momentum in the core (e.g., Burrows et al. 2007b).  Moreover, the
current generation of neutrino-powered core-collapse simulations
(e.g., Blondin \& 
Mezzacappa 2007; Burrows et al. 2007a; Marek \& Janka 2009) all
generate complex structures in the explosion without needing to invoke
jets.  In particular, these simulations commonly create
aspherical shocks dominated by low-order deformations that produce
unipolar or bipolar outflows that seem promising for generating the
core asphericities we infer from our polarization measurements.  It
remains to be seen whether these features persist in the upcoming
more-detailed three-dimensional models.

Unfortunately, most explosion models to date only simulate the
central core of the star and stop when an outward-going shock is
launched (or has failed to launch).  However, the shock continues to
evolve as it traverses the star and our observations can only probe
the final state of the core, which makes detailed comparisons of the
observed asphericities to those seen in the simulations difficult.
Only a few studies \citep{kifonidis06,hammer09} have followed the
aspherical shock wave as it propagates outward through the star.
Future work in this direction, along with radiative-transfer
calculations, is necessary to see if the existing models can reproduce
the observations. 

It is interesting to note that \ov\ showed the largest continuum
polarizations in our sample while it ejected the smallest amount of
nickel and possibly had the lowest-mass progenitor.  As mentioned
above, \citet{smartt09} quote a $^{56}$Ni mass of $0.003 
\pm 0.002$~M$_{\odot}$, a value about an order of magnitude lower than
in most SNe~IIP.  That \ov\ ejected a low $^{56}$Ni mass can also be
inferred by comparison of Figures~\ref{ov_lcfig} and \ref{aa_lcfig}.
\ov\ and \aaa\ had similar magnitudes on the plateau and \ov\ was at a
somewhat smaller distance \citep{smartt09}, yet it was a magnitude
fainter than \aaa\ at the beginning of the radioactive decay tail.
The progenitor mass limit for \ov\ is also very low, $M < 10$~M$_{\odot}$
\citep{smartt09,crockett09}.  In addition, the expansion velocities of
\ov\ are lower than those of the other SNe IIP (Figure~\ref{halfig}),
possibly indicating a low explosion energy.  A connection between the
low nickel mass, low progenitor mass, or low explosion energy and the
high continuum polarization would be exciting if confirmed in other
objects.

We also encourage the enlargement of the published SN~IIP
spectropolarimetry sample, particularly of objects
lacking significant ISP.  While observations at very late times are
likely rare in the extant unpublished datasets, it would be fruitful
to expand the number of objects plotted in Figure~\ref{allfig}.  Only
then will we be able to look for relationships between the core
asphericity and other supernova properties.  Trends with progenitor
mass, explosion energy, and $^{56}$Ni production will all be
interesting to examine and may provide constraints on the theoretical
uncertainties in the core-collapse supernova mechanism.

\acknowledgments 
We thank Douglas C. Leonard for supplying us with his SNe 1999em and
2004dj data and Ryan J. Foley, Mohan Ganeshalingam, Matthew
Moore, and Thea N. Steele for their assistance with some of the
observations.  Most of the data presented herein were obtained at the
W. M. Keck Observatory, which is operated as a scientific partnership
among the California Institute of Technology, the University of
California, and the National Aeronautics and Space Administration; it
was made possible by the generous financial support of the W. M. Keck
Foundation.  The authors wish to recognize and acknowledge the very
significant cultural role and reverence that the summit of Mauna Kea
has always had within the indigenous Hawaiian community; we are most
fortunate to have the opportunity to conduct observations from this
mountain.  We also would like to thank the expert assistance of the
Keck and Lick staffs in making these observations possible. A.V.F.'s
supernova group at U.C. Berkeley has been supported by NSF grants
AST-0607485 and AST-0908886, as well as by the TABASGO Foundation.
KAIT and its ongoing operation were made possible by donations from
Sun Microsystems, Inc., the Hewlett-Packard Company, AutoScope
Corporation, Lick Observatory, the NSF, the University of California,
the Sylvia \& Jim Katzman Foundation, and the TABASGO Foundation.

{\it Facilities:} \facility{Shane (Kast)}, \facility{Keck:I (LRIS)}

\end{document}